\documentclass[12pt]{article}
\usepackage[top=25truemm,bottom=27truemm,left=22truemm,right=22truemm]{geometry}

\usepackage{indentfirst}
\usepackage{graphicx}
\usepackage{float}
\usepackage{url}
\usepackage{siunitx}
\usepackage{array}
\usepackage{amsmath}
\usepackage{amssymb}
\usepackage{fancybox}
\usepackage{bm}
\usepackage{mathrsfs}
\usepackage{cancel}
\usepackage{physics}
\usepackage{color}
\usepackage{authblk}
\usepackage{appendix}
\usepackage{comment}
\usepackage{caption,latexsym,amsfonts,mathtools}
\usepackage{subcaption}
\usepackage[hidelinks]{hyperref}
\usepackage{multirow}

\numberwithin{equation}{section}

\newcommand{\beqn}{\begin{eqnarray}}
\newcommand{\eeqn}{\end{eqnarray}}
\newcommand{\MSE}{\mathcal M_{S|E}}
\renewcommand{\Tr}{\operatorname{Tr}}
\newcommand{\ord}{\operatorname{ord}}
\newcommand{\CGME}{C_{\rm GME}}

\begin{document}
\begin{titlepage}
\begin{flushright}
{YITP-26-119, KOBE-COSMO-26-10}
\end{flushright}

\vspace{50pt}

\begin{center}
{\large{\textbf{Contact order governs the onset of entanglement cascades}}}

\vspace{25pt}

{Sugumi Kanno$^{1,2,3}$ and Jiro Soda$^4$}
\end{center}

\vspace{20pt}

\shortstack[l]
{\hspace{1.8cm}\it {\small $^1$Department of Physics, Kyushu University, Fukuoka 819-0395, Japan} \\[5pt]
\hspace{1.6cm}\it {\small $^2$ Quantum and Spacetime Research Institute, Kyushu University}\\[5pt]
\hspace{1.6cm}\it {\small $^3$ Center for Gravitational Physics and Quantum Information,} \\[5pt]
\hspace{1.9cm}\it {\small Yukawa Institute for Theoretical Physics, Kyoto University,} \\[5pt]
\hspace{1.8cm}\it {\small $^4$Department of Physics, Kobe University, Kobe 657-8501, Japan}}

\vspace{2cm}
\begin{abstract}
Entangling rates describe a direct entangling channel, but give no information when that channel is forbidden. For finite-dimensional analytic pure-state dynamics, we show that the onset of genuine $(n+1)$-partite entanglement is determined by the contact order $m_\star$ between the physical trajectory and the $S|E$ product manifold. It is the first Taylor order that cannot be reproduced by any product curve, or equivalently the first nonvanishing order of the Fubini--Study distance from the product manifold. We derive a time-ordered recursion that removes curvature-induced kinematic terms and computes $m_\star$ from the Taylor coefficients of $H(t)$. This provides a geometric description of an entanglement cascade, in which new subsystems can join multipartite entanglement only after one or more interaction steps. A symmetry-protected three-qubit model realizes $m_\star=2$, and a three-mode bosonic example shows that genuine tripartite entanglement can arise even when the two newly formed reduced pairs remain separable.
\end{abstract}
\end{titlepage}

\tableofcontents

\section{Introduction}

In multipartite dynamics, an important question is not only how much entanglement is generated, but also when a new subsystem first becomes part of an already entangled state. Standard quantities such as entangling power, Hamiltonian entangling rates, and short-time purity loss answer this question when the interaction opens an entangling direction at first order~\cite{Zanardi:2000zz,Dur2001,VanAcoleyen2013, Yang2018,Cresswell2018}. However, they do not determine what happens when this direct first-order channel is forbidden.

This situation can arise because of symmetries, selection rules, or the need to pass through intermediate states. In such cases, the new subsystem may join the entangled state only after two or more interaction steps. A vanishing linear entangling rate then does not tell us whether entanglement begins at second order, third order, or does not appear locally at all. Related work has studied multipartite entanglement generation and redistribution~\cite{Samanta2026,Qiu2025,ZhangLiZhang2026}. Here we focus on the first perturbative order at which the new subsystem joins a genuinely multipartite entangled state.

To characterize this onset, we introduce the \emph{contact order} $m_\star$ between the physical trajectory and the manifold of states that remain product across the appended cut. We compare the physical time evolution with all possible local product curves. The first Taylor order that cannot be reproduced by any such product curve defines $m_\star$. The generic case is $m_\star=1$, while symmetry- or selection-rule-protected dynamics can give $m_\star\ge2$. Because this definition is based only on the physical trajectory and the product manifold, it does not depend on the choice of a particular entanglement measure.

At second and higher orders, one must also take into account the curvature of the product manifold. Even a trajectory that remains exactly product can have a higher derivative with a component normal to the initial tangent space. A naive projection of this normal component would therefore give a false signal of entanglement. We remove this purely kinematic contribution by comparing the physical trajectory with an osculating product curve. For time-dependent Hamiltonians, the Taylor expansion must in addition keep the correct Dyson time ordering.

This geometric viewpoint is natural for closed quantum systems, where entanglement is generated coherently, redistributed among subsystems, and can later recur. It is also potentially relevant to cosmology, where the universe is often treated as a globally closed quantum system~\cite{Kiefer1998,CampoParentani2008,Burgess2023}. We do not assume a particular cosmological Hamiltonian here. Instead, our aim is
to identify the local dynamical structure that determines when a new degree of freedom first joins multipartite entanglement.

The organization of the paper is as follows. In Section~2, we define the contact order and relate it to genuine multipartite entanglement, Schmidt coefficients, and reduced purity. In Section~3, we derive a practical expansion of the Schr\"odinger evolution and explain the osculating subtraction required at higher orders. In Section~4, we study a symmetry-protected three-qubit model in which the onset is quadratic and becomes linear when the protecting symmetry is weakly
broken. In Section~5, we examine a three-mode bosonic example and show how arbitrary finite contact orders can arise. Section~6 summarizes the main results and discusses a possible interpretation in terms of decoherence in cosmology.

\section{Contact order and entanglement cascade}
\label{sec:contact}

We begin by fixing the basic terminology. A bipartition $A|\bar A$ means that the full Hilbert space is written as ${\cal H}_A\otimes{\cal H}_{\bar A}$. A pure state is product across this cut if it can be written as $\ket{u}_A\ket{v}_{\bar A}$, and is entangled across the cut otherwise. A pure state of $N$ parties is genuinely $N$-partite entangled if it is entangled across every nontrivial bipartition.

For a pure bipartite state, the Schmidt decomposition gives
\begin{equation}
 \ket{\Psi}
 =\sum_{\alpha}s_\alpha
 \ket{u_\alpha}_S\ket{v_\alpha}_E,
 \qquad
 s_\alpha\geq0,
 \qquad
 \sum_\alpha s_\alpha^2=1.
 \label{eq:schmidtintro}
\end{equation}
The state is product exactly when only one Schmidt coefficient is nonzero. Thus, when the physical trajectory leaves the product manifold across the cut $S|E$, new Schmidt coefficients begin to open.

We call the successive inclusion of an initially uncorrelated subsystem into an already entangled state an \emph{entanglement cascade}. Let
\begin{equation}
 S=A_1\cdots A_n
\end{equation}
be initially in a genuinely $n$-partite entangled pure state $\ket{\psi_n}$. We append a subsystem $E$ in the state $\ket0$ and assume analytic time evolution,
\begin{equation}
 \begin{aligned}
 \ket{\Psi(t)}&=U(t,0)\ket{\psi_n,0},\\
 U(t,0)&={\cal T}\exp\!\left[-i\int_0^t ds\,H(s)\right].
 \end{aligned}
 \label{eq:evolution}
\end{equation}
At $t=0$, the state is product across the cut $S|E$, even though the subsystem $S$ may already contain genuine multipartite entanglement.

Let $\MSE$ denote the manifold of states that remain product across $S|E$. We write the physical ray as
\begin{equation}
 \gamma(t)=[\Psi(t)],
\end{equation}
and compare it with a product curve
\begin{equation}
 \eta(t)\subset\MSE
\end{equation}
that follows the physical trajectory as closely as possible near $t=0$. The \emph{contact order} $m_\star$ is the first Taylor order at which this matching fails. More formally,
\begin{equation}
 m_\star-1=
 \max\!\left\{
 r\geq0:\ \exists\,\eta(t)\subset\MSE,\quad
 j_0^r\eta=j_0^r\gamma
 \right\}.
 \label{eq:contact}
\end{equation}
Here $j_0^r$ denotes the Taylor data through order $r$. Thus, if the value and all Taylor coefficients through order $m_\star-1$ can be reproduced by a product curve, but the coefficient at order $m_\star$ cannot, then the new subsystem first enters an entangling direction at order $m_\star$. If the matching continues to all orders,
we define
\begin{equation}
 m_\star=\infty.
\end{equation}

In finite dimensions, the same quantity can be described by the Fubini--Study distance from the product manifold~\cite{BrodyHughston2001, WeiGoldbart2003},
\begin{equation}
 m_\star=
 \ord_{t=0}
 d_{\rm FS}\!\left(\gamma(t),\MSE\right).
 \label{eq:distance}
\end{equation}
Here $\ord_{t=0}$ denotes the lowest nonzero power in the short-time asymptotic expansion.
For two normalized rays $[\Psi]$ and $[\Phi]$, we use
\begin{equation}
 d_{\rm FS}([\Psi],[\Phi])
 =\arccos|\langle\Phi|\Psi\rangle|.
 \label{eq:fsdef}
\end{equation}
The distance to $\MSE$ is obtained by minimizing this expression over all product states. Eq.~\eqref{eq:distance} therefore identifies the first power of $t$ at which the physical state separates from its best local product approximation. Appendix~\ref{app:geometry} gives a coordinate derivation of this equivalence.

To describe the first transverse contribution explicitly, choose local Schmidt bases whose first basis vectors follow the dominant product component. In these bases, the coefficient matrix can be written as
\begin{equation}
 M(t)=
 \begin{pmatrix}
 s_0(t)&0\\
 0&X(t)
 \end{pmatrix},
 \qquad
 X(t)=\sum_{m\geq1}t^mX_m.
 \label{eq:schmidtnormal}
\end{equation}
The coefficient $s_0(t)$ is the dominant Schmidt amplitude, while the block $X(t)$ contains the components orthogonal to both dominant local directions. These are precisely the components that open new Schmidt channels across $S|E$.

We write
\begin{equation}
 X_m\longleftrightarrow\ket{\chi_m},
\end{equation}
for the corresponding normal-state vector, with
\begin{equation}
 \|\ket{\chi_m}\|=\|X_m\|_F,
\end{equation}
where $\|\cdot\|_F$ denotes the Frobenius norm. At the first unmatched order,
\begin{equation}
 X(t)
 =t^{m_\star}
 \left[X_{m_\star}+O(t)\right].
\end{equation}
Local changes of Schmidt basis do not change the first nonzero order, the rank, the singular values, or the norm
$\|X_{m_\star}\|_F$.

\begin{figure}[t]
 \centering
 \includegraphics[width=0.95\textwidth]{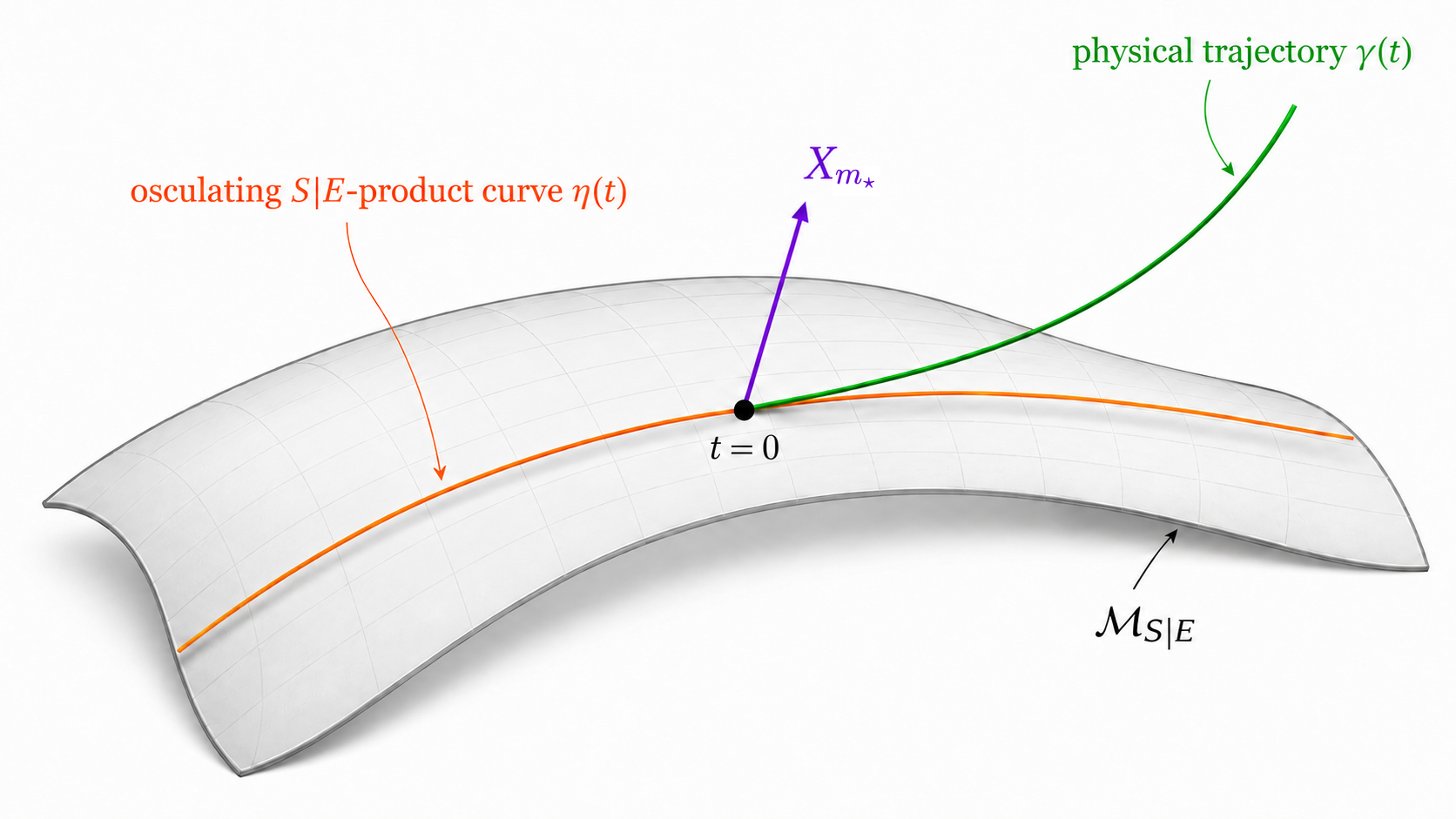}
 \caption{Contact geometry across the cut $S|E$. The physical trajectory $\gamma(t)$ and an osculating product curve $\eta(t)\subset\MSE$ agree through order $m_\star-1$. Their first unmatched coefficient is the transverse block $X_{m_\star}$. Since the product manifold is curved, higher derivatives of an exactly
 product trajectory can have normal components. These purely kinematic contributions must be removed before the physical contact order is identified.}
 \label{fig:contact}
\end{figure}

Figure~\ref{fig:contact} illustrates why higher derivatives cannot be interpreted by a simple normal projection. The product states form a curved manifold, so even a trajectory that remains exactly product can have a higher derivative with a normal component. The contact order is found only after the lower-order product motion has been matched and this kinematic bending has been removed.

\subsection{Local contact-order theorem}

We now state the local result that connects the geometric contact order to genuine multipartite entanglement. Assume that all party Hilbert spaces are finite dimensional and that the evolution in Eq.~\eqref{eq:evolution} is analytic near $t=0$. If $m_\star<\infty$, then there exists a $\delta>0$ such that $\ket{\Psi(t)}$ is genuinely $(n+1)$-partite entangled for every 
\begin{equation}
 0<t<\delta.
\end{equation}
If $m_\star=\infty$, the new cut $S|E$ remains product in a neighborhood of $t=0$.

The reason is simple. The initial state $\ket{\psi_n}$ is already genuinely $n$-partite entangled, so every bipartition inherited from the original subsystem $S$ is entangled at $t=0$. Because there are only finitely many such cuts and the evolution is continuous, these cuts remain entangled for sufficiently small $t$. The only cut that is initially product is $S|E$. Once this cut opens, every nontrivial bipartition is entangled, and the full state is therefore genuinely $(n+1)$-partite entangled. Conversely, if all transverse Taylor coefficients vanish, analyticity keeps the cut $S|E$ product locally. Appendix~\ref{app:theorem} gives the proof in detail.

Let $\sigma_\alpha$ denote the singular values of $X_{m_\star}$. The newly opened Schmidt coefficients then behave as
\begin{equation}
 s_\alpha(t)
 =t^{m_\star}\sigma_\alpha
 +O(t^{m_\star+1}).
 \label{eq:schmidt}
\end{equation}
Thus the contact order directly determines how quickly the new Schmidt channels open.

For the pure-state genuine multipartite concurrence~\cite{Ma2011},
\begin{equation}
 \CGME^{(n+1)}
 \equiv
 \min_{A|\bar A}
 \sqrt{2\left(1-\Tr\rho_A^2\right)},
 \label{eq:gmedef}
\end{equation}
the inherited cuts retain a finite amount of entanglement near $t=0$. Therefore, sufficiently close to the origin, the newly opened cut $S|E$ gives the minimum. We then obtain
\begin{equation}
 \CGME^{(n+1)}(t)
 =2t^{m_\star}\|X_{m_\star}\|_F
 +O(t^{m_\star+1}),
 \label{eq:gme}
\end{equation}
while the purity loss of the appended subsystem is
\begin{equation}
 P_E(t)\equiv1-\Tr\rho_E^2
 =2t^{2m_\star}\|X_{m_\star}\|_F^2
 +O(t^{2m_\star+1}).
 \label{eq:purity}
\end{equation}

The factors of two follow directly from the Schmidt coefficients. Let
\begin{equation}
 w(t)=\sum_{\alpha>0}s_\alpha^2(t)
\end{equation}
be the total weight of the newly opened Schmidt channels. Normalization gives
\begin{equation}
 s_0^2=1-w.
\end{equation}
Therefore
\begin{equation}
 1-\Tr\rho_E^2
 =1-\left[(1-w)^2+\sum_{\alpha>0}s_\alpha^4\right]
 =2w+O(w^2),
 \label{eq:purityquick}
\end{equation}
and
\begin{equation}
 \sqrt{2(1-\Tr\rho_E^2)}
 =2\sqrt{w}+O(w^{3/2}).
\end{equation}
Using
\begin{equation}
 s_\alpha(t)
 =t^{m_\star}\sigma_\alpha+\cdots
\end{equation}
then gives Eqs.~\eqref{eq:gme} and \eqref{eq:purity}.

Thus the concurrence and the purity loss measure the same opening of the Schmidt spectrum, but with different powers of the transverse amplitude. More generally, if a diagnostic depends at leading order on the transverse Schmidt amplitudes with homogeneous degree $p$, then its short-time behavior begins as
\begin{equation}
 t^{pm_\star}.
\end{equation}
The contact order $m_\star$ is therefore the measure-independent information that determines the onset.

\section{Hamiltonian expansion and osculating subtraction}
\label{sec:expansion}

We now show how the contact order can be obtained directly from the
Hamiltonian. We begin with the Schr\"odinger equation
\begin{equation}
 i\frac{\partial}{\partial t}\ket{\Psi(t)}
 =H(t)\ket{\Psi(t)} .
\end{equation}
Expanding the Hamiltonian around $t=0$ gives
\begin{equation}
 H(t)
 =H_0+t\dot H_0+\frac{1}{2}t^2\ddot H_0+\cdots .
\end{equation}
For the state, we use Taylor coefficients,
\begin{equation}
 \ket{\Psi(t)}
 =\ket{\Psi_0}
 +t\ket{v_1}
 +t^2\ket{v_2}
 +t^3\ket{v_3}
 +\cdots ,
 \qquad
 \ket{v_m}
 =\frac{1}{m!}
 \left.
 \frac{d^m}{dt^m}\ket{\Psi(t)}
 \right|_{t=0},
 \label{eq:vdef}
\end{equation}
where $m\geq1$ and
\begin{equation}
 \ket{\Psi_0}=\ket{\psi_n,0}.
\end{equation}
Substituting these expansions into the Schr\"odinger equation and
matching equal powers of $t$, we obtain
\begin{equation}
 \ket{v_1}
 =-iH_0\ket{\Psi_0},
 \qquad
 \ket{v_2}
 =-\frac12
 \left(H_0^2+i\dot H_0\right)
 \ket{\Psi_0}.
 \label{eq:v12}
\end{equation}
Appendix~\ref{app:dyson} gives the derivation through third order.

To separate tangent and entangling directions, define
\begin{equation}
 Q_S=1-\ket{\psi_n}\bra{\psi_n},
 \qquad
 Q_E=1-\ket0\bra0 ,
\end{equation}
and the normal-space projector
\begin{equation}
 P_N=Q_S\otimes Q_E .
\end{equation}
The operator $P_N$ selects the part of a state that is orthogonal to
both factors of the initial product state. Such a component cannot be
generated by changing only $\ket{\psi_n}$ or only $\ket0$, and
therefore points away from the product manifold.

At first order, the transverse coefficient is simply
\begin{equation}
 X_1\longleftrightarrow
 P_N\ket{v_1}
 =-iP_NH_0\ket{\psi_n,0}.
 \label{eq:x1}
\end{equation}
If this quantity is nonzero, the physical trajectory leaves the
product manifold at first order and
\begin{equation}
 m_\star=1.
\end{equation}

For an interaction of the form
\begin{equation}
 H_0=\sum_\mu A_\mu\otimes B_\mu,
\end{equation}
the norm $\|X_1\|_F^2$ is determined by the connected fluctuations
\begin{equation}
 \Delta A_\mu
 =A_\mu-\langle A_\mu\rangle,
 \qquad
 \Delta B_\mu
 =B_\mu-\langle B_\mu\rangle .
\end{equation}
This reproduces the familiar short-time entangling timescale. The
usual connected-fluctuation result is therefore the generic
$m_\star=1$ case.

When $X_1=0$, the first-order motion remains tangent to the product
manifold, and the second-order motion must be treated more carefully.
We can write
\begin{equation}
 \ket{v_1}
 =\ket{a_1,0}
 +\ket{\psi_n,b_1}
 +\lambda_1\ket{\psi_n,0},
 \label{eq:v1decomp}
\end{equation}
where
\begin{equation}
 \langle\psi_n|a_1\rangle=0,
 \qquad
 \langle0|b_1\rangle=0 .
\end{equation}
The last term changes only the norm and phase of the chosen
Hilbert-space representative. It is therefore longitudinal in
projective Hilbert space and can be removed by an appropriate choice
of representative.

Now consider a product curve with the same first-order motion,
\begin{align}
 \ket{\phi(t)}\ket{e(t)}
 ={}&
 \left(
 \ket{\psi_n}
 +t\ket{a_1}
 +\cdots
 \right)
 \left(
 \ket0
 +t\ket{b_1}
 +\cdots
 \right)
 \nonumber\\
 ={}&
 \ket{\psi_n,0}
 +t\left(
 \ket{a_1,0}
 +\ket{\psi_n,b_1}
 \right)
 +t^2\ket{a_1,b_1}
 +\cdots .
 \label{eq:productexpansion}
\end{align}
The term $\ket{a_1,b_1}$ is normal to the initial tangent space.
However, it is produced by a curve that remains exactly product for
all $t$. It therefore represents the curvature of the product
manifold rather than physical entanglement.

The physical second-order transverse coefficient is consequently
\begin{equation}
 X_2\longleftrightarrow
 P_N\!\left(
 \ket{v_2}-\ket{a_1,b_1}
 \right).
 \label{eq:x2}
\end{equation}
Appendix~\ref{app:osculating} gives the explicit projector
construction of $\ket{a_1}$ and $\ket{b_1}$ and derives this
second-order subtraction. Geometrically, the subtracted term describes the quadratic bending of
the product manifold. In a derivative expansion this bending is
described by the second fundamental form; our Taylor-coefficient
convention absorbs the corresponding numerical factor.

The need for this subtraction is already visible for a purely local
Hamiltonian,
\begin{equation}
 H_L=A\otimes1+1\otimes B,
\end{equation}
where the subscript $L$ denotes ``local''. Since $H_L$ contains no
interaction between the two subsystems, an initially product state
remains exactly product. Nevertheless, the projected second-order
coefficient can be nonzero,
\begin{equation}
 P_N\ket{v_2}
 =-Q_SA\ket{\psi_n}\otimes Q_EB\ket0
 =\ket{a_1,b_1}.
 \label{eq:falsepositive}
\end{equation}
A naive projection of $\ket{v_2}$ would therefore incorrectly
identify this purely kinematic bending as entanglement. The
subtraction in Eq.~\eqref{eq:x2} removes this false contribution.

As a nontrivial example with an interaction, consider two qubits
initially in $\ket{00}$ with the controlled Hamiltonian
\begin{align}
 H_{\rm ctrl}={}&
 \alpha\sigma_x\otimes1
 +\beta1\otimes\sigma_x
 +\lambda\sigma_x\otimes\ket1\bra1 ,
 \label{eq:ctrl}
\end{align}
where $\sigma_x$ is the Pauli $x$ matrix and ``ctrl'' refers to the
conditional last term. The first-order change is
\begin{equation}
 \ket{v_1}
 =-i\alpha\ket{10}
 -i\beta\ket{01}.
\end{equation}
This motion is tangent to the product manifold, and hence
\begin{equation}
 X_1=0.
\end{equation}

At second order,
\begin{equation}
 P_N\ket{v_2}
 =-\left(
 \alpha\beta+\frac{\beta\lambda}{2}
 \right)\ket{11}.
 \label{eq:controlledv2}
\end{equation}
The product motion itself contributes
\begin{equation}
 \ket{a_1,b_1}
 =-\alpha\beta\ket{11}.
\end{equation}
Subtracting this kinematic term gives
\begin{equation}
 X_2\longleftrightarrow
 -\frac{\beta\lambda}{2}\ket{11}.
\end{equation}
The detailed second-order calculation is given in
Appendix~\ref{app:osculating}.
The two-qubit concurrence therefore behaves as
\begin{equation}
 C_{\rm GME}^{(2)}(t)
 =|\beta\lambda|t^2+O(t^4).
\end{equation}
Thus, for $\beta\lambda\neq0$, the first physical transverse contribution appears at second order, and the example has
\begin{equation}
 m_\star=2.
\end{equation}

The same idea extends to arbitrary order. Write an osculating product
curve as
\begin{equation}
 \ket{\phi(t)}
 =\ket{\psi_n}
 +\sum_{r\geq1}t^r\ket{a_r},
 \qquad
 \ket{e(t)}
 =\ket0
 +\sum_{r\geq1}t^r\ket{b_r}.
\end{equation}
Suppose that the physical and product curves have already been matched
through order $m-1$. At order $m$, the lower-order product coefficients
generate the contribution
\begin{equation}
 \sum_{r=1}^{m-1}
 \ket{a_r,b_{m-r}}.
\end{equation}
We therefore define the residual
\begin{equation}
 \ket{R_m}
 =
 \ket{v_m}
 -
 \sum_{r=1}^{m-1}
 \ket{a_r,b_{m-r}} .
 \label{eq:Rm}
\end{equation}
Its tangent components determine the next coefficients
$\ket{a_m}$ and $\ket{b_m}$ of the osculating product curve, while its
normal component gives the physical transverse coefficient,
\begin{equation}
 X_m\longleftrightarrow
 P_N\ket{R_m}.
 \label{eq:Xmrec}
\end{equation}
Equivalently, defining
\begin{equation}
 K_m
 =
 P_N
 \sum_{r=1}^{m-1}
 \ket{a_r,b_{m-r}},
\end{equation}
we have
\begin{equation}
 X_m\longleftrightarrow
 P_N\ket{v_m}-K_m .
 \label{eq:Km}
\end{equation}
The first order for which
\begin{equation}
 X_m\neq0
\end{equation}
is the contact order $m_\star$. Although the individual coefficients
of the osculating product curve depend on the chosen local
coordinates, the first order at which the matching fails is
coordinate independent.

At higher orders, one must also keep the time ordering of the
Hamiltonian. For example, the third Taylor coefficient is
\begin{equation}
 \ket{v_3}
 =
 \left[
 \frac{i}{6}H_0^3
 -\frac{1}{6}H_0\dot H_0
 -\frac{1}{3}\dot H_0H_0
 -\frac{i}{6}\ddot H_0
 \right]
 \ket{\Psi_0}.
 \label{eq:v3}
\end{equation}
The two terms containing $H_0\dot H_0$ and $\dot H_0H_0$ have
different coefficients because the Hamiltonian at different times
need not commute. Thus the higher-order contact coefficients retain
information about Dyson time ordering that cannot be reconstructed
from powers of $H_0$ alone.

\section{Symmetry-protected three-qubit model}
\label{sec:qubit}

We now present a three-qubit example in which a two-step process is protected. For the subsystem $AB$, we use the Bell states
\begin{equation}
 \ket{\Phi^\pm}=\frac{\ket{00}\pm\ket{11}}{\sqrt2},\qquad
 \ket{\Psi^+}=\frac{\ket{01}+\ket{10}}{\sqrt2},
\end{equation}
and take
\begin{equation}
 \ket{\Psi_0}=\ket{\Phi^+}_{AB}\ket0_E.
\end{equation}
We consider the Hamiltonian
\begin{equation}
 H_{\rm loc}=\frac{\Delta}{2}(1-\sigma_x^A\sigma_x^B)
 +\frac{g_1}{2}(\sigma_z^A+\sigma_z^B)
 +\frac{g_2}{2}(\sigma_y^A+\sigma_y^B)\sigma_x^E.
 \label{eq:hloc}
\end{equation}
Here $\sigma_x$, $\sigma_y$, and $\sigma_z$ are the Pauli matrices, and the superscripts indicate the subsystem on which they act. The subscript ``loc'' emphasizes that the dynamics is generated by few-body couplings rather than by a direct transition from the initial state to the final entangled state. The $g_1$ term changes the state of $AB$ without affecting $E$, whereas the $g_2$ term can flip $E$ only after the required intermediate state of $AB$ has been generated.

The dynamics therefore closes in the basis
\begin{equation}
 \ket{e_0}=\ket{\Phi^+}\ket0,\qquad
 \ket{e_1}=\ket{\Phi^-}\ket0,\qquad
 \ket{e_2}=i\ket{\Psi^+}\ket1,
\end{equation}
and the Hamiltonian restricted to this invariant subspace is
\begin{equation}
 H_{\rm inv}=\begin{pmatrix}
 0&g_1&0\\
 g_1&\Delta&g_2\\
 0&g_2&0
 \end{pmatrix}.
 \label{eq:hinv}
\end{equation}
The absence of a direct matrix element between $e_0$ and $e_2$ makes the selection rule transparent. The first transition $e_0\to e_1$ remains in the manifold $\MSE$ because $E$ is still in $\ket0$. Only the second
transition $e_1\to e_2$ changes the state of $E$ and leaves the product manifold. The matrix in Eq.~\eqref{eq:hinv} is obtained by
acting with the Pauli operators in Eq.~\eqref{eq:hloc} on the three basis states; the individual actions are listed in Appendix~\ref{app:qubit}.

Expanding $e^{-iH_{\rm inv}t}\ket{e_0}=\ket{e_0}-itH_{\rm inv}\ket{e_0}-t^2H_{\rm inv}^2\ket{e_0}/2+\cdots$, we obtain
\begin{align}
 \ket{\Psi(t)}={}&\ket{e_0}-ig_1t\ket{e_1}\nonumber\\
 &-\frac{t^2}{2}\left(g_1^2\ket{e_0}+g_1\Delta\ket{e_1}
 +g_1g_2\ket{e_2}\right)+O(t^3).
 \label{eq:expansion3}
\end{align}
For $\Delta=0$, the absence of a direct $e_0\leftrightarrow e_2$ transition is protected by symmetry. Define
\begin{equation}
 \Gamma_{\rm full}=\sigma_x^A\sigma_x^B,\qquad
 Q=\sigma_z^A\sigma_z^B\sigma_z^E,
\end{equation}
and let $\Pi_{AB}$ denote the exchange $A\leftrightarrow B$. These operators satisfy
\begin{equation}
 \{\Gamma_{\rm full},H_{\rm loc}\}=0,\qquad
 [Q,H_{\rm loc}]=[\Pi_{AB},H_{\rm loc}]=0.
\end{equation}
The first relation is the chiral symmetry, while the other two show that the dynamics preserves the eigenvalue of $Q$ and the exchange symmetry of $A$ and $B$.

The three states $\ket{e_0}$, $\ket{e_1}$, and $\ket{e_2}$ are symmetric under $A\leftrightarrow B$ and satisfy
\begin{equation}
 Q\ket{e_j}=\ket{e_j},\qquad j=0,1,2.
\end{equation}
They therefore span the three-dimensional invariant subspace already introduced above. Under $\Gamma_{\rm full}$, the states $\ket{e_0}$
and $\ket{e_2}$ have the same eigenvalue, whereas $\ket{e_1}$ has the
opposite eigenvalue. Since the chiral symmetry allows the Hamiltonian
to connect only states with opposite eigenvalues of
$\Gamma_{\rm full}$, a direct $e_0\leftrightarrow e_2$ transition is forbidden.
For $g_1g_2\neq0$, the first nonzero transverse contribution therefore
appears at second order and is
\begin{equation}
 -\frac{g_1g_2}{2}t^2\ket{e_2}.
\end{equation}
Consequently,
\begin{equation}
 \CGME^{(3)}=|g_1g_2|t^2+O(t^4),\qquad
 P_E=\frac{g_1^2g_2^2}{2}t^4+O(t^6).
 \label{eq:protected}
\end{equation}
For $\Delta=0$, we write
\begin{equation}
 \ket{\Psi(t)}
 =c_0(t)\ket{e_0}+c_1(t)\ket{e_1}+c_2(t)\ket{e_2}.
\end{equation}
Solving the Schr\"odinger equation in this three-dimensional invariant subspace, we obtain the exact transverse amplitude
\begin{equation}
 c_2(t)=\frac{g_1g_2}{\Omega^2}
 \left[\cos(\Omega t)-1\right],\qquad
 \Omega=\sqrt{g_1^2+g_2^2}.
 \label{eq:c2exact}
\end{equation}
Expanding this result for short times gives
\begin{equation}
 c_2(t)=-\frac{g_1g_2}{2}t^2+O(t^4),
\end{equation}
in agreement with Eq.~\eqref{eq:expansion3}. Thus the quadratic onset of the transverse motion is already visible directly in the state amplitude, without introducing a particular entanglement measure.

A weak symmetry-breaking perturbation can restore a direct
first-order entangling channel. Consider
\begin{equation}
 \epsilon V,\qquad
 V=\frac12(\sigma_x^A+\sigma_x^B)\sigma_x^E .
\end{equation}
This perturbation preserves the $Q$ symmetry and the exchange symmetry $A\leftrightarrow B$, but breaks the chiral symmetry that forbids the direct transition. The leading behavior then becomes
\begin{equation}
 \CGME^{(3)}
 =2\left|\epsilon t+\frac{g_1g_2}{2}t^2+\cdots\right|.
 \label{eq:breaking}
\end{equation}
The crossover time $t_\times$, at which the linear and quadratic
terms become comparable, is therefore estimated as
\begin{equation}
 t_\times\sim\frac{2|\epsilon|}{|g_1g_2|}.
\end{equation}
For any nonzero $\epsilon$, the linear term dominates for
$t\ll t_\times$, so the true asymptotic contact order is
$m_\star=1$. For
$t_\times\ll t\ll\min(|g_1|^{-1},|g_2|^{-1})$, however,
the quadratic term can dominate and the symmetry-protected
$m_\star=2$ behavior remains visible in an intermediate quadratic
regime.

The contact order can also be extracted from the purity loss $P_E(t)$ defined in Eq.~\eqref{eq:purity}. Since
\begin{equation}
 P_E(t)\propto t^{2m_\star}
\end{equation}
at short times, we define the effective exponent
\begin{equation}
 m_{\rm eff}^{(P)}(t)
 =\frac12\frac{d\ln P_E(t)}{d\ln t}.
 \label{eq:meff}
\end{equation}
The factor $1/2$ appears because the purity loss is quadratic in the leading transverse amplitude.
For exact symmetry protection,
$m_{\rm eff}^{(P)}(t)\to m_\star=2$ as $t\to0$.
With weak symmetry breaking, the true asymptotic value becomes one,
while an intermediate plateau near two can remain visible before
higher-order and recurrence effects become important.

\begin{figure}[t]
 \centering
 \includegraphics[width=0.88\textwidth]{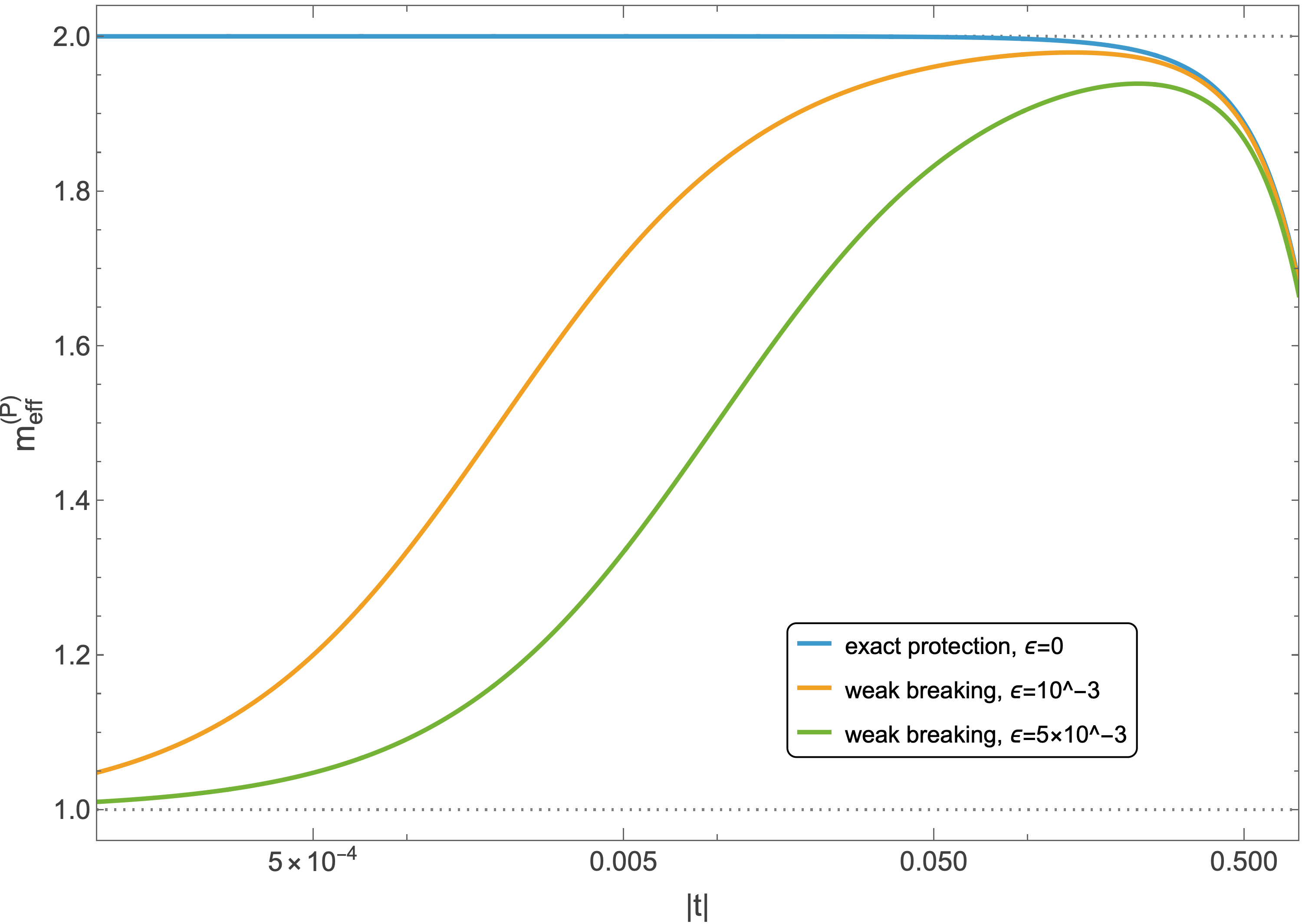}
\caption{Effective contact order $m_{\rm eff}^{(P)}$ extracted from the purity loss for $g_1=g_2=1$ and $\Delta=0$, with $\epsilon=0$, $10^{-3}$, and $5\times10^{-3}$. For exact protection ($\epsilon=0$), $m_{\rm eff}^{(P)}\to2$ as $t\to0$. Weak symmetry breaking changes the true short-time value to $1$, while an intermediate plateau near $2$ shows the remaining quadratic regime.}
 \label{fig:purity}
\end{figure}

Figure~\ref{fig:purity} shows this behavior numerically. For exact
symmetry protection, the effective exponent approaches $2$ at short
times. When the symmetry is weakly broken, it approaches $1$ as
$t\to0$, while a plateau near $2$ remains visible at intermediate
times.

\section{Bosonic example and higher contact-order hierarchy}
\label{sec:hierarchy}

We next consider a regular three-mode bosonic example that realizes
the generic first-order case. Let $a$, $b$, and $c$ denote bosonic
annihilation operators, and consider the interaction
\begin{equation}
 H_I=g\left(abc^\dagger+a^\dagger b^\dagger c\right).
 \label{eq:bosonHI}
\end{equation}
We take the initial state to be
\begin{equation}
 \ket{\Psi_0}
 =\sqrt{1-q^2}\sum_{m\ge0}q^m\ket{m,m,0},
 \qquad
 q=\tanh r,
 \label{eq:bosoninitial}
\end{equation}
where $r\ge0$ is the squeezing parameter. The modes $a$ and $b$ are
therefore initially in a two-mode squeezed state, while the mode $c$
starts in the vacuum and is uncorrelated with them. The mean occupation
number of each squeezed mode is
\begin{equation}
 \bar n=\sinh^2 r.
\end{equation}

At $t=0$, only the term $abc^\dagger$ acts nontrivially on the initial
state, because $c\ket0_c=0$. Hence the first Taylor coefficient is
\begin{equation}
 -iH_I\ket{\Psi_0}
 =-ig\sqrt{1-q^2}\sum_{m\ge1}m q^m
 \ket{m-1,m-1,1}.
 \label{eq:bosonfirst}
\end{equation}
After subtracting the component tangent to the initial product
manifold, the remaining transverse norm is
\begin{equation}
 \|X_1\|_F^2=g^2\bar n^2.
\end{equation}
The explicit sums are given in Appendix~\ref{app:boson}. Therefore
\begin{equation}
 \CGME^{(3)}
 =2\bar n|gt|+O(|gt|^3).
 \label{eq:boson}
\end{equation}
so that, for $g\neq0$ and nonzero squeezing $r>0$,
\begin{equation}
 m_\star=1.
\end{equation}

The physical meaning is simple. The interaction removes one quantum
from both $a$ and $b$ and creates one quantum in $c$, or performs the
reverse process. Since mode $c$ participates already in the first
nonzero transverse term, it joins the entangled state at first order.

The resulting tripartite entanglement is not simply a transfer of
pairwise entanglement from the pair $ab$ to the pairs $ac$ or $bc$.
Since the occupation-number difference $n_a-n_b$ is conserved and
vanishes initially, the occupations of $a$ and $b$ remain equal.
The state therefore has the form
\begin{equation}
 \ket{\Psi(t)}
 =\sum_{k,m}C_{km}\ket{k,k,m}.
 \label{eq:bosonstate}
\end{equation}
Tracing over mode $b$, define
\begin{equation}
 \ket{\varphi_k}=\sum_m C_{km}\ket m,
 \qquad
 p_k=\langle\varphi_k|\varphi_k\rangle.
\end{equation}
For $p_k\neq0$, let
\begin{equation}
 \ket{\widehat\varphi_k}
 =p_k^{-1/2}\ket{\varphi_k}.
\end{equation}
Then
\begin{equation}
 \rho_{ac}
 =\sum_k p_k\,\ket{k}\bra{k}_a
 \otimes
 \ket{\widehat\varphi_k}\bra{\widehat\varphi_k}_c .
\end{equation}
Since $p_k\geq0$ and $\sum_kp_k=1$, this is an explicit convex mixture of product states. Therefore $\rho_{ac}$ is separable. By the same argument, $\rho_{bc}$ is also separable.

Thus mode $c$ becomes part of a genuinely tripartite entangled state even though neither of the new reduced pairs $ac$ and $bc$ is
pairwise entangled. The cut $ab|c$ opens already at first order. The two inherited cuts
$a|bc$ and $b|ac$ are entangled at $t=0$ because the initial two-mode squeezed state is entangled across $a|b$, and they remain entangled for sufficiently small $t$ along this regular trajectory. Thus the global state is genuinely tripartite entangled for
sufficiently small nonzero times. At the same time, the reduced states $\rho_{ac}$ and $\rho_{bc}$ remain separable. This example therefore shows that the growth of genuine multipartite entanglement need not be understood as a simple redistribution of pairwise entanglement.

We use this bosonic model only as a regular trajectory with well-defined
Taylor coefficients. The finite-dimensional theorem proved above is
not assumed to extend automatically to a general infinite-dimensional
Hilbert space.

We can also realize any finite contact order in a finite-dimensional model. Let $\{\ket{s_j}\}_{j=0}^m$ be orthonormal states of $S$, with $\ket{s_0}$ genuinely $n$-partite entangled, and take nonzero couplings $g_j$. Define
\begin{equation}
 \ket{e_j}=\ket{s_j}\ket0_E
 \quad (0\le j<m),
 \qquad
 \ket{e_m}=\ket{s_m}\ket1_E.
\end{equation}
Consider the chain Hamiltonian
\begin{equation}
 H_m
 =\sum_{j=0}^{m-1}
 g_{j+1}
 \left(
 \ket{e_{j+1}}\bra{e_j}
 +\mathrm{H.c.}
 \right).
 \label{eq:chain}
\end{equation}
Starting from $\ket{e_0}$, the first $m-1$ steps keep the subsystem
$E$ in $\ket0_E$. These contributions can therefore be reproduced by
a product curve. The first amplitude that changes the state of $E$
appears at order $m$,
\begin{equation}
 \frac{(-it)^m}{m!}
 \left(\prod_{j=1}^{m}g_j\right)
 \ket{e_m}
 +O(t^{m+1}).
 \label{eq:chainamp}
\end{equation}
Hence
\begin{equation}
 X_1=\cdots=X_{m-1}=0,
 \qquad
 X_m\neq0,
\end{equation}
and therefore
\begin{equation}
 m_\star=m.
\end{equation}

This shows that every finite contact order can occur as the first
physically allowed transverse process. For a sparse and
time-independent interaction graph, $m_\star$ may coincide with the
length of the shortest nonzero path to a state in which the appended
subsystem has changed. In a more general system, however, interference
between different Dyson paths and the osculating subtraction can
modify this simple graph interpretation.

In a finite closed system, the redistribution of entanglement remains
coherent and can recur. If the appended subsystem is not observed, the
same unitary process appears locally as the purity loss introduced in
Eq.~\eqref{eq:purity}. Irreversible decoherence requires additional
physics, such as a continuum of modes, dephasing, coarse graining, or
a thermodynamic limit. The contact order nevertheless identifies the
first unitary channel through which information can leave the
subsystem being followed.

The value of $m_\star$ depends on both the initial state and the chosen
description of the system. The same Hamiltonian can therefore have
different contact orders for different initial states, because the
tangent and normal spaces of the product manifold depend on the
reference state. Likewise, if an intermediate degree of freedom is
integrated out, a microscopic two-step process can appear as a direct
effective coupling. The invariance of $m_\star$ should therefore be
understood after the resolved degrees of freedom and the physical
trajectory have been fixed.

Finally, our local theorem assumes analytic, finite-dimensional
pure-state dynamics. Finite dimensionality ensures a smooth local
product manifold and only finitely many inherited bipartitions. Mixed
states require the more complicated geometry of separable density
operators. For nonanalytic driving, all Taylor coefficients may vanish
even though the trajectory does not remain locally inside the product
set. These extensions are beyond the scope of the present work.
\section{Conclusion}

We have shown that the onset of genuine multipartite entanglement is not always determined by a linear entangling rate. Instead, it is controlled
by the contact order $m_\star$ between the physical time-ordered trajectory and the product manifold associated with the appended cut. The integer $m_\star$ is the first Taylor order that cannot be reproduced by any product curve. It is therefore a local geometric property of the trajectory and does not depend on the choice of a particular entanglement measure.

The same contact order determines the leading short-time behavior of several physical quantities. The new Schmidt coefficients and the genuine multipartite concurrence scale as $t^{m_\star}$, while the purity loss of the appended subsystem scales as $t^{2m_\star}$. The short-time logarithmic slope of the purity loss therefore provides a direct way to extract $m_\star$ from the appended subsystem alone.

At second and higher orders, the curvature of the product manifold must be treated carefully. Even an exactly product trajectory can have a nonzero normal component in its higher derivatives. The osculating subtraction removes this purely kinematic contribution. Together with the time-ordered expansion of the Hamiltonian, it gives a practical procedure for finding the first physical transverse coefficient.

Our three-qubit model provides a simple example with $m_\star=2$. The symmetry forbids a direct transition, so the appended qubit can join the entangled state only through a two-step process. A weak symmetry-breaking term restores a first-order channel and changes the true short-time order to $m_\star=1$, although the quadratic behavior can remain visible over an intermediate time range. The three-mode bosonic example gives the generic case $m_\star=1$ and shows that genuine tripartite entanglement can appear even when the two newly formed reduced pairs remain separable.

The first nonzero transverse coefficient contains more information than the contact order alone. The data
\begin{equation}
 \left(m_\star,\operatorname{rank}X_{m_\star},
 \{\sigma_\alpha\}\right)
\end{equation}
describe the order at which the new entangling channel opens, the number of Schmidt channels that open at that order, and their relative strengths. The norm $\|X_{m_\star}\|_F$ gives the overall strength of the leading transverse process.

If new subsystems are added successively, these contact orders define an entanglement cascade,
\begin{equation}
 {\cal E}_2\xrightarrow{m_1}{\cal E}_3\xrightarrow{m_2}{\cal E}_4
 \xrightarrow{m_3}\cdots ,
 \label{eq:hierarchy}
\end{equation}
where ${\cal E}_n$ denotes the set of genuinely $n$-partite entangled pure states for the chosen factorization. The usual connected-fluctuation result corresponds to the generic case $m_\star=1$. Higher contact orders describe symmetry- or selection-rule-protected entanglement growth that cannot be detected from a linear entangling rate alone.

Our results suggest a new way to think about decoherence in cosmology. Decoherence is often described as entanglement between a system and its environment. Our entanglement cascade shows that the process can be more structured: quantum correlations can spread step by step, bringing more degrees of freedom into a genuinely multipartite entangled state. From this point of view, decoherence need not begin with a single direct system--environment interaction. Instead, entanglement can spread step by step through many degrees of freedom, forming an entanglement cascade. The contact order $m_\star$ tells us when this cascade can first begin, even when the lowest-order entangling channel is forbidden. We therefore propose that entanglement cascades provide a new geometric picture of how decoherence may develop in cosmological systems.

\appendix
\section{Detailed derivations}
\label{app:derivations}

This appendix gives the intermediate algebra needed to derive the main results directly from the initial state and Hamiltonian, using only ordinary perturbation theory and the Schmidt decomposition.

\subsection{Local coordinates near the product manifold}
\label{app:geometry}

Let $\ket{\Psi_0}=\ket{\psi}\ket0$ be normalized. Choose orthonormal bases $\{\ket\psi,\ket{s_i}\}$ for $S$ and $\{\ket0,\ket{e_a}\}$ for $E$. A nearby state can be written as
\begin{equation}
 \ket{\Psi}=c_{00}\ket{\psi,0}+\sum_i c_{i0}\ket{s_i,0}
 +\sum_a c_{0a}\ket{\psi,e_a}+\sum_{ia}c_{ia}\ket{s_i,e_a}.
 \label{eq:localcoords}
\end{equation}
The first three terms are the value plus tangent directions of the product manifold at $\ket{\Psi_0}$; the last block is normal. Indeed, a nearby product state has the form
\begin{align}
 &(\ket\psi+\sum_i x_i\ket{s_i})(\ket0+\sum_a y_a\ket{e_a})\nonumber\\
 &\quad=\ket{\psi,0}+\sum_i x_i\ket{s_i,0}+\sum_a y_a\ket{\psi,e_a}
 +\sum_{ia}x_i y_a\ket{s_i,e_a}.
 \label{eq:localproduct}
\end{align}
Thus the normal coefficients of a product state are not independent: to lowest nontrivial order they are the products $x_i y_a$. This elementary identity is the local origin of the osculating subtraction.

In the coefficient-matrix language, Eq.~\eqref{eq:localcoords} is a block matrix. Local unitary rotations can eliminate the off-diagonal tangent blocks order by order, leaving the Schmidt-normal form of Eq.~\eqref{eq:schmidtnormal}. If the first nonzero transverse block is $X(t)=t^mX_m+O(t^{m+1})$, the closest product ray differs from the physical ray by a vector of norm $t^m\|X_m\|_F+O(t^{m+1})$. For two nearby normalized rays, $d_{\rm FS}=\|\delta\Psi_\perp\|+O(\|\delta\Psi_\perp\|^2)$. Hence
\begin{equation}
 d_{\rm FS}(\gamma(t),\MSE)=t^m\|X_m\|_F+O(t^{m+1}),
\end{equation}
which proves Eq.~\eqref{eq:distance} locally.

\subsection{Why opening the new cut implies genuine multipartite entanglement}
\label{app:theorem}

At $t=0$, the state $\ket{\psi_n}\ket0_E$ is product only across the appended cut $S|E$. Every bipartition inherited from the original parties cuts the genuinely entangled state $\ket{\psi_n}$ and therefore has positive linear entropy. Denote these finitely many values by
\begin{equation}
 L_A(0)=1-\Tr\rho_A^2(0)>0,
\end{equation}
where $A|\bar A$ runs over the inherited cuts. Because there are finitely many cuts,
\begin{equation}
 L_{\min}=\min_{A|\bar A\,\mathrm{inherited}}L_A(0)>0.
\end{equation}
The density matrices and their purities depend continuously on $t$, so there exists $\delta_1>0$ such that $L_A(t)>L_{\min}/2$ for every inherited cut and $t<\delta_1$. If $m_\star<\infty$, then Eq.~\eqref{eq:schmidt} gives a nonzero second Schmidt coefficient across $S|E$ for sufficiently small nonzero $t$; hence there is $\delta_2>0$ for which that cut is entangled whenever $0<t<\delta_2$. Taking $\delta=\min(\delta_1,\delta_2)$ proves that every bipartition is entangled in this punctured neighborhood. For a pure state this is exactly genuine $(n+1)$-partite entanglement.

If every transverse Taylor coefficient vanishes, analyticity implies $X(t)=0$ in a neighborhood of the origin. The state then remains rank one in Schmidt form across $S|E$, so the appended party does not join locally.

For completeness, let $s_0^2=1-w$ and $w=\sum_{\alpha>0}s_\alpha^2$. Then
\begin{align}
 \Tr\rho_E^2&=(1-w)^2+\sum_{\alpha>0}s_\alpha^4,\\
 1-\Tr\rho_E^2&=2w-w^2-\sum_{\alpha>0}s_\alpha^4=2w+O(w^2).
\end{align}
Since $w=t^{2m_\star}\sum_\alpha\sigma_\alpha^2+O(t^{2m_\star+1})$, Eq.~\eqref{eq:purity} follows. Taking the square root gives Eq.~\eqref{eq:gme} because the appended cut is the minimizing cut close to $t=0$.

\subsection{Time-ordered Taylor coefficients}
\label{app:dyson}

Write
\begin{equation}
 H(t)=\sum_{r\ge0}\frac{t^r}{r!}H_r,
 \qquad
 H_r=\left.\frac{d^rH}{dt^r}\right|_{t=0},
 \qquad
 \ket{\Psi(t)}=\sum_{m\ge0}t^m\ket{v_m},
\end{equation}
with
\begin{equation}
 \ket{v_0}=\ket{\Psi_0}.
\end{equation}
Substituting these expansions into the Schr\"odinger equation
$i\partial_t\ket{\Psi}=H\ket{\Psi}$ and matching equal powers of
$t$, we obtain
\begin{equation}
 i(m+1)\ket{v_{m+1}}
 =\sum_{r=0}^{m}\frac{H_r}{r!}\ket{v_{m-r}}.
 \label{eq:recursionv}
\end{equation}
The first two steps give Eq.~\eqref{eq:v12}. At the next step,
\begin{align}
 \ket{v_3}
 &=-\frac{i}{3}\left(
 H_0\ket{v_2}
 +H_1\ket{v_1}
 +\frac12H_2\ket{v_0}
 \right)\nonumber\\
 &=\left[
 \frac{i}{6}H_0^3
 -\frac16H_0H_1
 -\frac13H_1H_0
 -\frac{i}{6}H_2
 \right]\ket{\Psi_0},
\end{align}
where
\begin{equation}
 H_1=\dot H_0,\qquad H_2=\ddot H_0.
\end{equation}
This is Eq.~\eqref{eq:v3}. The unequal coefficients of
$H_0H_1$ and $H_1H_0$ are the local imprint of Dyson time ordering.
\subsection{Second-order osculating subtraction}
\label{app:osculating}

Apply the projectors $Q_S$ and $Q_E$ to $\ket{v_1}$ and define
\begin{equation}
 \ket{a_1}=Q_S\otimes\bra0\,\ket{v_1},\qquad
 \ket{b_1}=\bra\psi\otimes Q_E\,\ket{v_1}.
\end{equation}
When $P_N\ket{v_1}=0$, these vectors reproduce all nonlongitudinal first-order motion. Equation~\eqref{eq:productexpansion} then shows that the product curve contributes $P_N\ket{a_1,b_1}=\ket{a_1,b_1}$ at second order. Subtracting it yields Eq.~\eqref{eq:x2}.

For the purely local Hamiltonian $H_L=A\otimes1+1\otimes B$,
\begin{equation}
 \ket{a_1}=-iQ_SA\ket\psi,\qquad \ket{b_1}=-iQ_EB\ket0.
\end{equation}
The cross term in $H_L^2$ is $2A\otimes B$, so
\begin{equation}
 P_N\ket{v_2}=-Q_SA\ket\psi\otimes Q_EB\ket0=\ket{a_1,b_1}.
\end{equation}
Hence $X_2=0$, as it must for an exactly factorized evolution $e^{-iAt}\ket\psi\otimes e^{-iBt}\ket0$.

For Eq.~\eqref{eq:ctrl}, $H_{\rm ctrl}\ket{00}=\alpha\ket{10}+\beta\ket{01}$. Thus
\begin{equation}
 \ket{a_1}=-i\alpha\ket1,\qquad \ket{b_1}=-i\beta\ket1,
\end{equation}
so $\ket{a_1,b_1}=-\alpha\beta\ket{11}$. Acting once more with the Hamiltonian and retaining the $\ket{11}$ component gives
\begin{equation}
 \bra{11}H_{\rm ctrl}^2\ket{00}=2\alpha\beta+\beta\lambda.
\end{equation}
Therefore $P_N\ket{v_2}=-(\alpha\beta+\beta\lambda/2)\ket{11}$, and subtraction leaves $X_2=-(\beta\lambda/2)\ket{11}$. A two-qubit pure state $a\ket{00}+b\ket{01}+c\ket{10}+d\ket{11}$ has concurrence $C=2|ad-bc|$; inserting the short-time amplitudes gives $C(t)=|\beta\lambda|t^2+O(t^4)$.

\subsection{Three-qubit model}
\label{app:qubit}

The following Bell-state identities are sufficient to reproduce Eq.~\eqref{eq:hinv}:
\begin{align}
 (\sigma_z^A+\sigma_z^B)\ket{\Phi^+}&=2\ket{\Phi^-},&
 (\sigma_z^A+\sigma_z^B)\ket{\Phi^-}&=2\ket{\Phi^+},\\
 (\sigma_y^A+\sigma_y^B)\ket{\Phi^-}&=2i\ket{\Psi^+},&
 (\sigma_y^A+\sigma_y^B)\ket{\Psi^+}&=-2i\ket{\Phi^-},\\
 (1-\sigma_x^A\sigma_x^B)\ket{\Phi^-}&=2\ket{\Phi^-},&
 (1-\sigma_x^A\sigma_x^B)\ket{\Phi^+}&=0.
\end{align}
Together with $\sigma_x^E\ket0=\ket1$ and $\sigma_x^E\ket1=\ket0$, these relations give
\begin{equation}
H_{\rm loc}\ket{e_0}=g_1\ket{e_1},\quad
H_{\rm loc}\ket{e_1}
=g_1\ket{e_0}+\Delta\ket{e_1}+g_2\ket{e_2},\quad
H_{\rm loc}\ket{e_2}=g_2\ket{e_1},
\end{equation}
which is Eq.~\eqref{eq:hinv}. It follows immediately that
\begin{equation}
 H_{\rm loc}^2\ket{e_0}
 =g_1^2\ket{e_0}+g_1\Delta\ket{e_1}+g_1g_2\ket{e_2}.
\end{equation}
This reproduces Eq.~\eqref{eq:expansion3}.

For $\Delta=0$, the amplitudes satisfy
\begin{equation}
 i\dot c_0=g_1c_1,\qquad i\dot c_1=g_1c_0+g_2c_2,\qquad i\dot c_2=g_2c_1,
\end{equation}
with $(c_0,c_1,c_2)=(1,0,0)$ at $t=0$. Subtracting $(g_2/g_1)$ times the first equation from the third shows that $c_2-(g_2/g_1)c_0=-g_2/g_1$ is constant. Eliminating $c_1$ then gives $\ddot c_0+\Omega^2c_0=g_2^2$, with $\Omega^2=g_1^2+g_2^2$. The solution is
\begin{equation}
 c_0=\frac{g_2^2+g_1^2\cos\Omega t}{\Omega^2},\qquad
 c_2=\frac{g_1g_2}{\Omega^2}(\cos\Omega t-1),
\end{equation}
which establishes Eq.~\eqref{eq:c2exact}.

The weak-breaking operator satisfies
\begin{equation}
 V\ket{e_0}=-i\ket{e_2}.
\end{equation}
Therefore the first-order correction to the state is
\begin{equation}
 -i\epsilon t V\ket{e_0}
 =-\epsilon t\ket{e_2}.
\end{equation}
Together with the symmetry-protected second-order contribution $-(g_1g_2/2)t^2\ket{e_2}$, this gives the modulus appearing in Eq.~\eqref{eq:breaking}. Balancing the linear and quadratic terms gives
\begin{equation}
 t_\times\sim\frac{2|\epsilon|}{|g_1g_2|}.
\end{equation} 

\subsection{Bosonic first-order norm and pairwise separability}
\label{app:boson}

Let
\begin{equation}
 \ket{\psi_r}_{ab}
 =\sqrt{1-q^2}\sum_{m\ge0}q^m\ket{m,m},
 \qquad q=\tanh r ,
\end{equation}
be the two-mode squeezed vacuum state of modes $a$ and $b$.
The initial state is
\begin{equation}
 \ket{\Psi_0}=\ket{\psi_r}_{ab}\ket0_c .
\end{equation}
Since $c\ket0=0$,
\begin{equation}
 H_I\ket{\Psi_0}=g\,ab\ket{\psi_r}\otimes\ket1_c.
\end{equation}
The mean of $ab$ in the two-mode squeezed vacuum is
\begin{equation}
 \langle ab\rangle=(1-q^2)\sum_{m\ge1}m q^{2m-1}=\frac{q}{1-q^2}=\sqrt{\bar n(\bar n+1)}.
\end{equation}
The norm of $ab\ket{\psi_r}$ is
\begin{equation}
 \langle a^\dagger a\,b^\dagger b\rangle
 =(1-q^2)\sum_{m\ge0}m^2q^{2m}=\bar n+2\bar n^2.
\end{equation}
The component parallel to the initial $ab$ vector corresponds to a local change of mode $c$ and must be removed by $Q_S$. Hence
\begin{align}
 \|X_1\|_F^2
 &=g^2\left(\langle a^\dagger a\,b^\dagger b\rangle-|\langle ab\rangle|^2\right)\nonumber\\
 &=g^2\left(\bar n+2\bar n^2-\bar n(\bar n+1)\right)
 =g^2\bar n^2,
\end{align}
which proves Eq.~\eqref{eq:boson}.

For the separability statement, conservation of $n_a-n_b$ implies equal occupations of $a$ and $b$, so the state can be expanded as $\sum_{k,m}C_{km}\ket{k,k,m}$. Tracing over $b$ uses $\langle k'|k\rangle=\delta_{kk'}$ and gives
\begin{align}
 \rho_{ac}
 &=\sum_{k,m,m'}C_{km}C_{km'}^*\ket{k,m}\bra{k,m'}\\
 &=\sum_k p_k\,\ket{k}\bra{k}_a\otimes\ket{\widehat\varphi_k}\bra{\widehat\varphi_k}_c,
\end{align}
where $p_k=\sum_m|C_{km}|^2$ and $\ket{\widehat\varphi_k}=p_k^{-1/2}\sum_mC_{km}\ket m$ when $p_k\neq0$. Since $p_k\ge0$ and $\sum_kp_k=1$, this is an explicit convex decomposition into product states. Thus $\rho_{ac}$ is separable; the same argument applies to $\rho_{bc}$.

\subsection{Arbitrary contact order from a chain}
\label{app:chain}

For the Hamiltonian in Eq.~\eqref{eq:chain}, the only way to reach $\ket{e_m}$ from $\ket{e_0}$ in fewer than $m$ applications of $H_m$ would be to skip a link, which the Hamiltonian does not contain. Therefore
\begin{equation}
 \bra{e_m}H_m^r\ket{e_0}=0\quad(r<m),\qquad
 \bra{e_m}H_m^m\ket{e_0}=\prod_{j=1}^m g_j.
\end{equation}
Expanding $e^{-iH_mt}$ gives Eq.~\eqref{eq:chainamp}. Since $E$ remains in $\ket0$ for all basis states $e_0,\ldots,e_{m-1}$, all lower-order coefficients belong to a product curve across $S|E$. The first coefficient with $E=\ket1$ therefore occurs at order $m$, proving $m_\star=m$.

\vspace{1cm}
\section*{Acknowledgments}
S.\ K. was supported by the Japan Society for the Promotion of Science (JSPS) KAKENHI Grant No. JP24K21548. J.\ S. was in part supported by JSPS KAKENHI Grants No. JP23K22491, No. JP24K21548, and No. JP25H02186.

\bibliographystyle{unsrt}
\bibliography{references}

@article{Zanardi:2000zz,
    author = "Zanardi, Paolo and Zalka, Christof and Faoro, Lara",
    title = "{Entangling power of quantum evolutions}",
    eprint = "quant-ph/0005031",
    archivePrefix = "arXiv",
    doi = "10.1103/PhysRevA.62.030301",
    journal = "Phys. Rev. A",
    volume = "62",
    pages = "030301",
    year = "2000"
}

@article{Dur2001,
  author={D\"ur, Wolfgang and Vidal, Guifr\'e and Cirac, J. Ignacio and Linden, Noah and Popescu, Sandu},
  title={Entanglement capabilities of nonlocal Hamiltonians},
  journal={Phys. Rev. Lett.}, volume={87}, pages={137901}, year={2001},
  doi={10.1103/PhysRevLett.87.137901}}

@article{VanAcoleyen2013,
  author={Van Acoleyen, Karel and Mari\"en, Micha\"el and Verstraete, Frank},
  title={Entanglement rates and area laws},
  journal={Phys. Rev. Lett.}, volume={111}, pages={170501}, year={2013},
  doi={10.1103/PhysRevLett.111.170501}}

@article{Yang2018,
  author={Yang, I-Sheng}, title={The entanglement timescale},
  journal={Phys. Rev. D}, volume={97}, pages={066008}, year={2018},
  doi={10.1103/PhysRevD.97.066008}}

@article{Cresswell2018,
  author={Cresswell, Jesse C.}, title={Universal entanglement timescale for R\'enyi entropies},
  journal={Phys. Rev. A}, volume={97}, pages={022317}, year={2018},
  doi={10.1103/PhysRevA.97.022317}}

@article{Samanta2026,
  author={Samanta, Mrinmoy and Mondal, Sudipta and Hazra, Samir Kumar and Sen (De), Aditi},
  title={Hierarchies among genuine multipartite entangling capabilities of quantum gates},
  journal={Phys. Rev. A}, volume={113}, pages={012612}, year={2026}, doi={10.1103/vhvd-jbnk}}

@article{Qiu2025,
  author={Qiu, Xinyu and Song, Zhiwei and Chen, Lin},
  title={Multipartite entangling power by von Neumann entropy},
  journal={Phys. Rev. A}, volume={111}, pages={022407}, year={2025}, doi={10.1103/PhysRevA.111.022407}}

@article{ZhangLiZhang2026,
  author={Zhang, Chun-Yue and Li, Zi-Xiang and Zhang, Shi-Xin},
  title={Entanglement Growth from Entangled States: A Unified Perspective on Entanglement Generation and Transport},
  journal={Phys. Rev. Lett.}, volume={137}, pages={020404}, year={2026}, doi={10.1103/xkh7-gdqm}}

@article{Kiefer1998,
  author={Kiefer, Claus and Polarski, David and Starobinsky, Alexei A.},
  title={Quantum-to-classical transition for fluctuations in the early universe},
  journal={Int. J. Mod. Phys. D}, volume={7}, pages={455}, year={1998}}

@article{CampoParentani2008,
  author={Campo, David and Parentani, Renaud},
  title={Decoherence and entropy of primordial fluctuations. I. Formalism and interpretation},
  journal={Phys. Rev. D}, volume={78}, pages={065044}, year={2008}, doi={10.1103/PhysRevD.78.065044}}

@article{Burgess2023,
  author={Burgess, C. P. and Holman, R. and Kaplanek, Greg and Martin, J\'er\^ome and Vennin, Vincent},
  title={Minimal decoherence from inflation},
  journal={J. Cosmol. Astropart. Phys.}, volume={2023}, number={07}, pages={022}, year={2023},
  doi={10.1088/1475-7516/2023/07/022}}

@article{BrodyHughston2001,
  author={Brody, Dorje C. and Hughston, Lane P.}, title={Geometric quantum mechanics},
  journal={J. Geom. Phys.}, volume={38}, pages={19}, year={2001}, doi={10.1016/S0393-0440(00)00052-8}}

@article{WeiGoldbart2003,
  author={Wei, Tzu-Chieh and Goldbart, Paul M.},
  title={Geometric measure of entanglement and applications to bipartite and multipartite quantum states},
  journal={Phys. Rev. A}, volume={68}, pages={042307}, year={2003}, doi={10.1103/PhysRevA.68.042307}}

@article{Ma2011,
  author={Ma, Zhi-Hao and Chen, Zhi-Hua and Chen, Jing-Ling and Spengler, Christoph and Gabriel, Andreas and Huber, Marcus},
  title={Measure of genuine multipartite entanglement with computable lower bounds},
  journal={Phys. Rev. A}, volume={83}, pages={062325}, year={2011}, doi={10.1103/PhysRevA.83.062325}}

\end{document}